\documentstyle[preprint,aps,epsf]{revtex}

\newcommand{\beq}{\begin{equation}}
\newcommand{\eeq}{\end{equation}}
\newcommand{\bea}{\begin{eqnarray}}
\newcommand{\eea}{\end{eqnarray}}
\newcommand{\bef}{\begin{figure}}
\newcommand{\enf}{\end{figure}}
\newcommand{\ba}{\begin{array}}
\newcommand{\ea}{\end{array}}

\newcommand{\ket}[1]{| #1 \rangle}
\newcommand{\bra}[1]{\langle #1 |}
\newcommand{\proj}[1]{\ket{#1}\! \bra{#1}}

\newcommand{\inner}[2]{ \langle #1 | #2 \rangle}
\newcommand{\melement}[2]{ \langle #1 | #2 | #1 \rangle}
\newtheorem{theorem}{Theorem}
\newtheorem{lemma}{Lemma}
\newtheorem{definition}{Definition}

\def\cd{{\cal D}}
\def\ce{{\cal E}}
\def\ci{{\cal I}}
\def\ch{{\cal H}}
\def\ct{{\cal T}}

\def\tr{{\rm Tr}}
\def\qbs{Q_E}
\def\qbdsw{Q_P}
\def\zerozero{\proj{0^C}}
\def\eig{\phi_{\rm max}}
\def\chin{\chi^{\otimes N}}

\begin{document}
\title{The quantum capacity is properly defined without encodings}

\author{Howard Barnum$^{(1)}$, 
John A. Smolin$^{(2)}$, 
Barbara M. Terhal$^{(3)}$} 
\address{$^{(1)}$Hampshire College 
and Institute for Science and Interdisciplinary Studies, Amherst, 
MA, 01022, USA.\\
Email: {\tt hbarnum@hampshire.edu} \\ 
$^{(2)}$IBM Research Division, T.J. Watson Research Center, Yorktown Heights, New York 10598, USA.\\
 Email: {\tt smolin@watson.ibm.com} \\
$^{(3)}$Faculteit WINS, Universiteit van Amsterdam\\ 
Valckenierstraat 65, 1018 XE Amsterdam and\\
Centrum voor Wiskunde en Informatica,
Kruislaan 413, 1098 SJ Amsterdam, The Netherlands.\\
Email: {\tt terhal@phys.uva.nl} \\
}

\maketitle

\begin{abstract}

We show that no source encoding is needed in the definition of the
capacity of a quantum channel for carrying quantum information.  This
allows us to use the coherent information maximized over all sources
and and block sizes, but not encodings, to bound the quantum capacity.
We perform an explicit calculation of this maximum coherent information
for the quantum erasure channel and apply the bound in order 
find the erasure channel's capacity without relying on an unproven 
assumption as in an earlier paper.

\end{abstract}
\vspace{.1 in}
\pacs{PACS: 03.65.Bz, 89.70.+c}
\vspace{.2 in}
\narrowtext

\section{Introduction}

In recent years the field of quantum information theory has emerged.
One of the central issues in this field is the concept of 
quantum channel capacity. Several papers have discussed the 
capacity of noisy quantum channels to carry quantum 
information\cite{schum1,schum2,Lloyd,BDSW,Barnum,erasure}.  
Unfortunately defining and calculating the quantum capacity has turned out 
to be difficult, 
because of the specific (and sometimes odd) features of quantum 
information.  Various other types of capacities of quantum
channels have also been defined, such as the capacity of
a quantum channel to carry classical information \cite{holevo,SW}, the 
capacities of quantum channels to carry quantum information with the
assistance of classical side-channels \cite{BDSW}, and a
capacity based on a quantum analogue of the Shannon mutual information
\cite{adami}.  Here we will concentrate on just one type of quantum
capacity.

Barnum, Nielsen and Schumacher \cite{Barnum} have given
a definition of quantum capacity $\qbs(\chi)$ of a channel $\chi$
in terms of the entanglement fidelity and the 
von Neumann entropy $S(\rho)\equiv-{\rm Tr}\; \rho \log \rho$ of the 
source's density matrix $\rho$.  

The entanglement fidelity of a density matrix $\rho$
relative to a linear trace-preserving completely positive map
$\ce$ \cite{superoperator} is defined as 
\beq
F_e(\rho,\ce)=\melement{\eta}{\;(\ci \otimes \ce)(\proj{\eta})\;}
\label{fentanglement}
\eeq 
where $\ket{\eta}$ 
is any purification of $\rho$. 
A {\em purification} \cite{jozsa} of any density matrix $\rho$ 
in a Hilbert space $\cal H$ is any pure state $\ket{\eta}$ 
in a tensor product space  $\ch_A \otimes \ch_B$ such that 
${\rm Tr}_A \proj{\eta} = \rho$.  In Equation (\ref{fentanglement})
the identity operates on the purification space $\ch_A$ and 
$\ce$ operates on $\ch_B$.  Note that 
$F_e(\rho,\ce)$ is independent of the choice of purification
\cite{schum1}.

\begin{definition}
\label{qbsdef}
The entanglement capacity $\qbs$ of a channel $\chi$ is 
\beq
\label{capacityBS}
\qbs(\chi) \equiv 
\sup\{q:   \forall_{\epsilon > 0} 
\,  
\exists_{\ce,\cd,\rho,N}: 
\frac{S(\rho)}{N}=q \;{\rm and}\; 	   
      F_e(\rho,\cd \circ \chi^{\otimes N} \circ \ce) > 1-\epsilon\}\;.
\eeq
\end{definition}

That is, roughly, $\qbs$ is the highest entropy per use of the 
channel which can be sent reliably using block coding.
Here the density operator $\rho$ is on a block of $N$ copies of the
input Hilbert space, and the encoding and decoding
operations $\ce$ and $\cd$ (which are linear trace-preserving completely
positive maps) act on such block density
operators.  The definition requires that arbitrarily high entanglement
fidelities may be achieved, possibly by going to larger and larger 
block size $N$.  It does not, however, require that arbitrarily high 
fidelity be achievable for some fixed block size $N$.  It is immediately 
apparent from the definition that one may bound this 
capacity below by some constant $r$ (for rate)
by exhibiting a sequence (in $N$)
of source density operators and coding schemes such that 
the entropy of the source operators goes to $r$ and the 
entanglement fidelity of the operators under the total operation
goes to $1$ with large $N$.  We will say such a sequence of 
triplets $(\rho, \ce, \cd)$ {\em achieves the rate} $r$.

The definition of $\qbs$ uses the entropy of the 
source $\rho$ as a measure of the information that is sent through
the channel rather than the entropy of the output signal 
$({\cal D}\circ \chi^{\otimes N} \circ \ce)(\rho)$.  
One might argue that since capacity is about sending entropy 
to the channel output one should consider a definition 
$Q^{\rm out}$ in which the entropy of the output signal appears 
in place of the entropy of the input $\rho$ as in $\qbs$.
But in general, as the decoding process $\cd$ need not be unitary
(and indeed cannot be if it is to extract the noise from the output
signal) it can map the signal onto an arbitrarily large Hilbert space, and 
the output entropy can become unboundedly large. This implies that 
$Q^{\rm out}$ is not a good measure of the total amount of information 
that is sent through the channel.  The problem is that
for any pure state there exist density matrices of high fidelity
relative to that pure state which have arbitrarily high entropy.
Consider the density matrix 
$\rho=(1-\epsilon) \proj{\psi} + {\epsilon\over n} 
\sum_{i=1}^n \proj{i}$ with the $\ket{i}$s an orthonormal set of vectors
orthogonal to $\ket{\psi}$.  
This density matrix has entropy $H_2(\epsilon)+ \epsilon \log n$  
and fidelity $1-\epsilon$ relative to $\ket{\psi}$ for any $\epsilon$ and
any $n$.
($H_2(\epsilon)= -\epsilon \log_2(\epsilon) -(1-\epsilon) 
\log_2(1-\epsilon')$ is the binary entropy function.)



Another quantity which has been of interest is the coherent information
\cite{schum1,Lloyd}.
\begin{definition}
The coherent information of a density matrix $\rho$ and a linear trace-preserving
completely positive map $\ce$ is 	
\begin{equation}
I_c(\rho,\ce)=S(\ce(\rho))-S_{\rm env}(\rho,\ce),
\end{equation}
where $S_{\rm env}(\rho,\ce)$ is the final entropy of an initially
pure environment implementing $\chi$ \cite{superoperator}. 
\end{definition}

Barnum, Nielsen and Schumacher \cite{Barnum} have shown that
\begin{equation}
\qbs \leq I_{\rm max} \equiv \sup_N\; \max_{\rm \rho,\ce} \frac{I_c(\rho, 
\chi^{\otimes N} \circ \ce)}{N}
\end{equation}
It has been conjectured \cite{schum1,Lloyd,Barnum} that this 
bound is an equality.

Notice that the definition of $\qbs$ includes a supremum over
encodings. This is required to give a 
most general definition of a channel capacity, but it is surprising 
from a physical point of view. Any unitary encoding of a source
is equivalent to using a different source and since the supremum also
includes the source, the unitary encoding could be left out. 
The coherent information, due to the failure of the pipelining 
inequality, {\em can} increase
by using non-unitary encoding (see \cite{Barnum}), which suggests
the necessity of the supremum over non-unitary encodings in 
the capacity definition.  But a non-unitary encoding intuitively 
corresponds to adding noise to the signal, which seems 
unlikely to improve the quality of the output signal.  This illustrates 
the complexity of the issue. In this paper we 
resolve this matter by showing that the supremum over
encodings can be omitted from the definition of capacity, though we
do not know if the maximization over encodings can be omitted
from $I_{\rm max}$.  

Another issue is the continuity of the quantum channel capacity in the
parameters of channel $\chi$. It is not known whether $\qbs$ or
$\qbdsw$ are continuous.  It was stated in \cite{erasure} that the
capacity of the erasure channel is $Q=\max\{0,1-2p\}$. This result was
derived by bounding the capacity both from below and from above with
$\max\{0,1-2p\}$.  The derivation of the upper bound however assumed
the capacity to be continuous as a function of $p$, which has not 
been proved.
We will use the results in this paper
to prove the capacity in an alternative way, thus resolving the 
continuity question for the erasure channel.  A similar proof
of the capacity of the erasure channel was carried out independently
(and first) by Cerf \cite{cerf} using a different definition of the 
quantum channel capacity.


In this paper we prove the following:
\begin{itemize}
\item The maximization over encodings $\ce$ in the definition of $\qbs$
is not necessary. In other words we find that
\beq
\qbs=\qbs^{\mbox{no encoding}}.
\eeq
where $\qbs^{\mbox{no encoding}}$ is defined exactly as is $\qbs$,
except without the encoding map $\ce$.
over encodings. See Sec. \ref{sec1}. 
\item The quantum capacity $\qbs$ is bounded from above by the maximum
 coherent information without {\em source encoding}
\beq
\qbs \le \lim_{N \rightarrow \infty} \max_{\rm \rho} \frac{I_c(\rho, 
\chi^{\otimes N})}{N}. 
\eeq
See Sec. \ref{sec3}.
\item The quantum capacity of the erasure channel \cite{erasure}
is given by $\qbs=\qbdsw=\max\{1-2p,0\}$ as in \cite{erasure}.
See Sec. \ref{sec3}. 
\end{itemize}

\section{$\qbs$ is well defined without source encoding}
\label{sec1}

Consider a situation where the sequence of
triplets $(\rho,\cd,\ce)$ achieves 
$\qbs$ and the $\ce$'s may be non-unitary.
We will show that there exists another sequence of triplet 
$(\rho',\ct \circ \cd,\ci)$ that achieves the capacity $\qbs$,
where $\ct$ is an additional decoding step. We thus 
replace the non-unitary {\em encoding} by a not-necessarily-unitary 
{\em decoding}.  We will do this by showing that for any triplet
$(\rho,\cd,\ce)$ with a given entropy and with a given entanglement fidelity
when used with the channel $\chi$, there exists another triplet
$(\rho',\ct \circ \cd,\ci)$ whose entropy and entanglement fidelity
are both close to those of the original triplet.

\subsection{Preliminaries}
We will need the following two lemmas:

\begin{lemma}
\label{enttheorem}
Given two bipartite pure states $\ket{\psi}$ and $\ket{\phi}$ in
a Hilbert space $\ch=\ch_A \otimes \ch_B$ with 
$|\inner{\psi}{\phi}|^2 \ge 1-\epsilon$ then
\beq
|\;S(\tr_A \proj{\psi})-S(\tr_A \proj{\phi})\;|\; \le 2 \sqrt{\epsilon} \log d + 1 
\eeq
for all $\epsilon <  \frac{1}{36}$ where $d$ is the dimension of $\ch_B$.
\end{lemma}

\noindent{\bf Proof} :
We will use an inequality from Fannes \cite{Ohya83a} involving 
the $L_1$ norm.
The $L_1$ norm of an operator $A$, indicated by $||A||$, is defined by
\beq
||A|| \equiv {\rm Tr} |A| \equiv {\rm Tr} \sqrt{A^\dagger A}\ . 
\eeq
We also define the function $\eta(x) = - x \log{x}$ and 
let $\rho_1,\rho_2$ be density matrices in $\ch_B$. 
We than have from \cite{Ohya83a}
(when $||\rho_1 - \rho_2 || <  \frac{1}{3}$) 
\beq \label{eqtn: L1normcontinuity} |S(\rho_1) -
S(\rho_2)| \le ||\rho_1 - \rho_2|| \log{d} + \eta(||\rho_1 -
 \rho_2||)\ .
\eeq 
For our purposes, we may note that for 
$x < \frac{1}{3}$, $\eta(x) < \frac{\log 3}{3} < 1$, and use the weaker
inequality 
\beq \label{eqtn: L1normcontinuityweak} 
|S(\rho_1) - S(\rho_2)| \le \log d ||\rho_1 - \rho_2|| + 1 \; .  
\eeq 
For two commuting density matrices $\rho_1$ and $\rho_2$ we 
have $|| \rho_1-\rho_2|| = \sum_i |\lambda_i^{(1)}-\lambda_i^{(2)}|$
with $\lambda_i^{(1,2)}$ the eigenvalues of density matrices 
$\rho_1$, $\rho_2$ respectively.
Since the entropy difference is invariant under independent unitary
rotations of each density matrix, 
\beq |S(\rho_1) - S(\rho_2)|
\le \log d \;\sum_i |\lambda_i^{(1)}-\lambda_i^{(2)}| + 1\;, 
\eeq 
where we have rearranged the eigenvalues in order of size.
It is known \cite{Kailath} that 
\beq 
\sum_i |\lambda_i^{(1)}-\lambda_i^{(2)}| \le 2\sqrt{1 - B(\lambda^{(1)},\lambda^{(2)})}\;, 
\eeq 
where $B$ is the Bhattacharyya-Wootters overlap \cite{BWOverlap}, defined by
 \beq B(\lambda^{(1)},\lambda^{(2)}) \equiv 
\left(\sum_i \sqrt{\lambda_i^{(1)}\lambda_i^{(2)}}\right)^2\;.  
\eeq 
The fidelity between two density matrices $\rho_1,\rho_2$ can be defined as the maximum
inner product between all purifications $\ket{\zeta_1},\ket{\zeta_2}$ of
$\rho_1$ and $\rho_2$: 
\beq
F(\rho_1,\rho_2)=\max_{\ket{\zeta_1},\ket{\zeta_2}} |\inner{\zeta_1}{\zeta_2}|^2\ .
\eeq
Since, given the eigenvalues of two density operators, the fidelity is
maximized by choosing their eigenvectors to be the same (assigned to
eigenvalues in order of size) 
\beq
B(\lambda^{(1)},\lambda^{(2)}) \ge F(\rho_1,\rho_2)\ .  
\eeq 
Hence 
\beq 
\label{eqtn: shmoo} 
|S(\rho_1) - S(\rho_2)|
\le 2 \sqrt{1 - F(\rho_1, \rho_2)} \log{d} + 1 
\eeq 
when 
\beq 2
\sqrt{1 - F(\rho_1, \rho_2)}) < \frac{1}{3}\;.  
\eeq 
And by the definition of $F(\rho_1,\rho_2)$ (which includes
a maximization) we have that 
\beq |S(\rho_1) - S(\rho_2)| \le 2 \sqrt{1 - |\langle \psi | \phi \rangle|^2}
 \log{d} + 1
\eeq 
where $\ket{\psi}$ and $\ket{\phi}$ are purifications of $\rho_1$ and $\rho_2$,
i.e. $\tr_A \proj{\psi}=\rho_1$ and $\tr_A \proj{\phi}=\rho_2$.
This holds whenever 
$F(\rho_1,\rho_2) > 1 - \frac{1}{36}$ which is certainly true whenever 
$|\langle \psi | \phi \rangle |^2 > 1 - \frac{1}{36}$.  $\Box$

\begin{lemma}
\label{enttheorem2}
Given a bipartite pure state $\ket{\phi}$ and density matrix
$\rho$ in Hilbert space $\ch=\ch_A \otimes \ch_B$ with
$\melement{\phi}{\rho} \ge 1-\epsilon$ and $\epsilon < \frac{1}{72}$ then 
\beq
|S(\tr_A \proj{\phi}) - S(\tr_A \;\rho)|  \le 2 \sqrt{2 \epsilon} \log \dim \ch_B  + 2 \label{part1}
\eeq
and similarly for system $B$, and thus 
\beq
|S(\tr_A\; \rho)- S(\tr_B \;\rho)|  \le 4 \sqrt{2 \epsilon} \log 
\max\{\dim \ch_A,\dim \ch_B\} + 4\label{part3}\ .
\eeq
\end{lemma}

\noindent{\bf Proof} : 
We can write 
\beq
\rho=(1-\epsilon') \proj{\phi_{\rm max}} + \epsilon' \rho'
\label{epsi}
\eeq
with $\epsilon' \le \epsilon$. This is obtained by diagonalizing
$\rho$ and noting that the largest eigenvalue of a density matrix is
always no smaller than the largest diagonal element of the matrix
\cite{kyfan}. $\ket{\eig}$ is the eigenvector of $\rho$ corresponding
to its largest eigenvalue.

Here is the plan for the proof. We will first bound 
$|S(\tr_A \; \rho)-S(\tr_A \proj{\eig})|$. Then we will argue that
$\ket{\eig}$ has high fidelity with respect to $\ket{\phi}$ and 
use Lemma \ref{enttheorem} 
to bound $|S(\tr_A \proj{\phi})-S(\tr_A \proj{\eig})|$ which will 
finally give us a bound on $|S(\tr_A \proj{\phi})-S(\tr_A \;\rho)|$.

Recall the property of the entropy \cite{wehrl}
\beq
\sum_i \lambda_i S(\rho_i) \leq S\left(\sum_i \lambda_i \rho_i\right) \leq \sum_i \lambda_i S(\rho_i)- \sum_i 
\lambda_i \log \lambda_i. 
\label{entprop}
\eeq
with $\sum_i \lambda_i=1$ and $\rho_i$ are density matrices.

Taking the partial trace of (\ref{epsi}) and using (\ref{entprop})
one can derive that 
\bea
\epsilon'S(\tr_A \;\rho')-\epsilon' S(\tr_A\; \proj{\eig})) \leq 
S(\tr_A\; \rho)-S(\tr_A\; \proj{\eig}) \nonumber \\
 \leq  \epsilon' S(\tr_A\; \rho')-\epsilon'
S(\tr_A \;\proj{\eig})+H_2(\epsilon'),
\eea
and thus
\beq
|S(\tr_A\;\rho)-S(\tr_A \proj{\eig})| \leq  \epsilon \log \dim \ch_B +1
\label{plan1}
\eeq .  

To prove that $\ket{\phi}$ and $\ket{\eig}$ have high fidelity we use 
Eq. (\ref{epsi}) and $\melement{\phi}{\rho}\ge 1-\epsilon$ to write
\beq
\melement{\phi}{\rho}=(1-\epsilon') |\inner{\phi}{\eig}|^2
+ \epsilon' \melement{\phi}{\rho'} \ge 1-\epsilon\ .
\eeq
The inner product $\melement{\phi}{\rho'}$ is no bigger than one
and  $\epsilon'\le\epsilon$ so we can rearrange things to get
\beq
|\inner{\phi}{\eig}|^2 \ge 1-2\epsilon\ .
\eeq
Thus, by Lemma \ref{enttheorem} we can bound
\beq
|S(\tr_A \proj{\phi})-S(\tr_A \proj{\eig})| \le \sqrt{2 \epsilon} \log\dim\ch_B + 1 \;.  
\label{plan2}
\eeq

Therefore we find, with (\ref{plan1}) and (\ref{plan2}), 
\beq
|S(\tr_A \proj{\phi} )-S(\tr_A\; \rho)| \leq 2 \sqrt{2 \epsilon} \log 
\dim\ch_B  +2.
\eeq
Finally, using $\tr_A \proj{\phi} =\tr_B \proj{\phi}$ for all pure states
and (\ref{part1}), we immediately have (\ref{part3}). $\Box$.

\vspace{.1 in}

\subsection{The main theorem}
\begin{theorem}\label{theorem: main}
Suppose $\rho$ a density operator on a Hilbert space
$\ch_A$ and $\ce, \cd$ linear trace-preserving completely positive
operations such that
\beq
F_e(\rho,\cd \circ \chin \circ \ce) \ge 1-\epsilon\ .
\eeq
Then there exist a density operator $\rho'$ and a linear trace-preserving
completely positive operation $\ct$ such that
\beq
F_e(\rho' ,{\cal T} \circ \cd \circ \chi^{\otimes N}) \ge 1-2\epsilon \; 
\eeq
and
\beq
|S(\rho)-S(\rho')| \le 2 \sqrt{2 \epsilon} \log \dim \ch_A + 2 \ .
\eeq
\end{theorem}

The proof consists of two parts. First we show if there exists 
a source $\rho$ that has high entanglement fidelity using some 
encoding $\ce$ and decoding $\cd$, 
we can always find another source $\rho'$ which 
has a high entanglement fidelity as well, but has additional decoding instead
of encoding. Secondly we show that this new source $\rho'$ has very nearly 
the same von Neumann entropy as $\rho$.  

\begin{figure}[htbp]
\epsfxsize=14.0cm
\leavevmode
\epsfbox{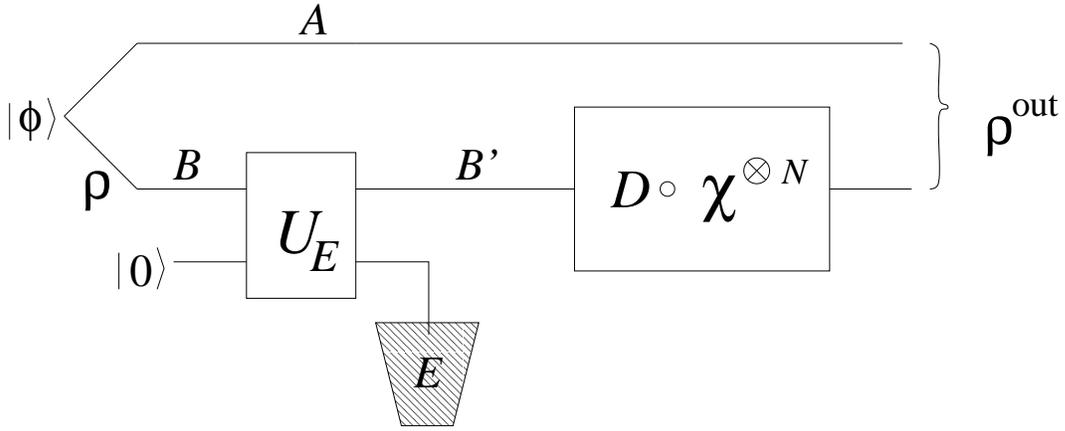}
\vspace{.1 in}
\caption{A general encoding-channel-decoding system.  $U_E$ is the
unitary operation of the encoder (the associated environment $E$ makes
the whole action of the encoder non-unitary in general).  
\label{fig1}}
\end{figure}

Let $\ket{\phi}$ be a purification
of $\rho$ in Hilbert space $\ch_A \otimes \ch_B$. See Fig.~\ref{fig1}. 
Any linear trace-preserving completely positive map, including non-unitary operations, can be written as a unitary operator
which operates on the original system along with an ancillary system
(often referred to as an environment), as in Fig.~\ref{fig1}.  Thus,
for the case of the non-unitary encoder, some quantum system 
$E$ which is in general entangled with the $AB'$ system
will remain in the encoder.  Since this system is not to be sent
through the channel it may be measured in an orthogonal basis giving
result $i$ with probability $p_i$ and leaving the $AB'$ system in a pure state
$\ket{\psi_i}$.  After the channel operates on the $B'$ system and the
decoding process is performed, one is left with 
$\rho_i^{\rm out}=(\ci_A \otimes (\cd \circ \chin)_B)(\proj{\psi_i})$.
(To simplify the notation we will hereafter write 
$\ci_A \otimes (\cd \circ \chin)_B$ as $\cd \circ \chin$.)
The whole encoding-channel-decoding process results in a high
entanglement fidelity so that
\beq
F_e(\rho,\cd \circ \chin \circ \ce)=\sum_i p_i \,\melement{\phi}{
(\cd \circ \chin)(\proj{\psi_i})} \ge 1-\epsilon\ .
\label{eq37a}
\eeq
For at least one value of $i$ it must be that
\beq
\melement{\phi}{(\cd \circ \chin)(\proj{\psi_i})} \ge 1-\epsilon\ .
\label{eq37}
\eeq
Thus, the unitary encoder that simply takes $\ket{\phi}$ and rotates it to
$\ket{\psi_i}$ is sufficient to achieve a high entanglement fidelity.
Hereafter the $i$ subscript will be dropped from $\ket{\psi_i}$ and $\rho_i^{\rm  out}$.

\begin{figure}[htbp]
\epsfxsize=14.0cm
\leavevmode
\epsfbox{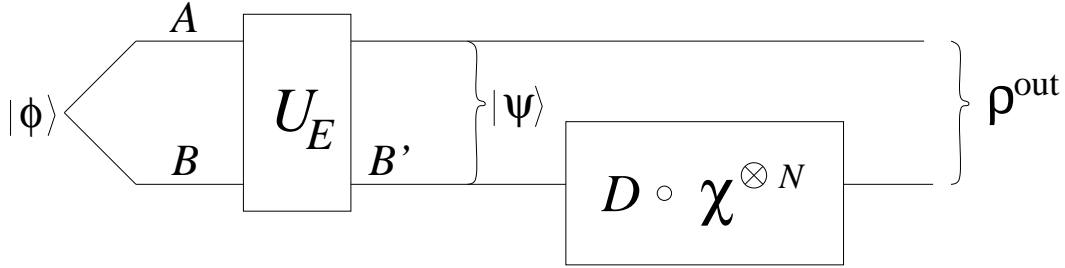}
\caption{The channel with unitary encoder acting on both the 
$A$ and $B$ system.  
\label{fig2}}
\end{figure}

We are now, however, left in the odd situation in which the unitary
encoder operates on both the $B$ and the $A$ systems.  We have thus so
far only traded non-unitarity for this odd form of unitarity.  This
situation is shown in Fig.~\ref{fig2}.  We will show that instead of
using $\ket{\phi}$ as input, we can use the unencoded $\ket{\psi}$ as
input if we do an additional decoding step.  The following Lemma will
be of use.

\begin{lemma}
\label{lemma1}
Given a density matrix $\rho$ in Hilbert space $\ch_A \otimes \ch_B$ 
then there exists a purification $\ket{\Psi}$ of $\tr_B\; \rho$
into Hilbert space $\ch_A \otimes \ch_B \otimes \ch_C$ with
$\dim \ch_C=\dim \ch_A + 1$ and
\beq
\melement{\Psi}{(\rho\otimes \zerozero) } = \lambda_{\rm max}^2
\eeq
where $\lambda_{\rm max}$ is the largest eigenvalue of $\rho$.
\end{lemma}

\noindent{\bf Proof} : We can write $\rho\otimes \zerozero$ as
\beq
\rho\otimes\zerozero=\lambda_{\rm max} \proj{\eig}\otimes\proj{0^C} + 
(1-\lambda_{\rm max}) \rho'\otimes \zerozero
\label{anequation}
\eeq
where $\ket{\eig}$ is the eigenvector of $\rho$ corresponding to 
$\lambda_{\rm max}$.
Take
\beq
\ket{\Psi}= \sqrt{\lambda_{\rm max}} \ket{\eig}\otimes\ket{0^C} + 
\sqrt{1-\lambda_{\rm max}}\sum_{i=1}^{\dim \ch_A} \sqrt{\mu_i} \ket{i^A}
\otimes \ket{0^B} \otimes \ket{i^C}
\eeq
where $\ket{i^A}$ and $\mu_i$ are the eigenvectors and eigenvalues 
of $\tr_B\; \rho$ and $\inner{0^C}{i^C}=0$. Thus 
$\melement{\Psi}{(\rho\otimes \zerozero) } = \lambda_{\rm max}^2$. $\Box$

\vspace{.5 in}

Since $\melement{\phi}{\rho^{\rm out}}\ge 1-\epsilon$ we have (as in Eq. (\ref{epsi}))
$\lambda_{\rm max}  \ge 1-\epsilon$.
Take $\ket{\Psi}$ also purifying $\tr_B(\rho^{\rm out})$ as in the
lemma.  Then 
\beq
\melement{\Psi}{(\rho^{\rm out} \otimes \zerozero)} \ge 
(1-\epsilon)^2\ge 1-2\epsilon\ .
\label{here}
\eeq

Since $\ket{\psi}$ purifies $\tr_B( \rho^{\rm out})$ so does 
$\ket{\psi_0}\equiv\ket{\psi}\otimes \ket{0^C}$.  
As $\ket{\Psi}$ and $\ket{\psi_0}$ both purify 
$\tr_B(\rho) \otimes \zerozero$, they are related by a 
unitary transformation $U=\ci_A \otimes U_{\rm BC}$ 
acting only on $\ch_B$ and $\ch_C$ \cite{purify}
\beq
U \ket{\Psi}=\ket{\psi_0}\ .
\eeq
Substituting this into (\ref{here}) and writing $\rho_0^{\rm out}\equiv \rho^
{\rm out}\otimes\zerozero$, we obtain
\beq
\melement{\psi_0}{U\rho_0 U^\dag}
\ge 1-2\epsilon \ .
\label{withc}
\eeq

We will now rid ourselves of the $C$ system.  As
\beq
\melement{\psi}{\tr_C U\rho_0^{\rm out} U^\dag} = 
\melement{\psi \otimes 0^C }{U\rho_0 U^\dag} + \sum_{i \ne 0}
\melement{\psi \otimes i^C }{U\rho_0 U^\dag}
\eeq
with $\melement{\psi \otimes i^C }{U\rho_0^{\rm out} U^\dag} \ge 0$ since
$U\rho_0^{\rm out} U^\dag$ is a density matrix, we can rewrite (\ref{withc}) as
\beq
\melement{\psi}{\tr_C U\rho_0^{\rm out} U^\dag } \ge 1-2\epsilon\ .
\eeq
Let us define $\ct (\rho^{\rm out})$ be the linear trace-preserving completely positive map
implemented by appending a $\ket{0^C}$ state to $\rho^{\rm out}$, rotating
using $U$ and then tracing out the $C$ system.  What we have done
is replaced $\ket{\phi}$ with $\ket{\psi}$ and added the decoding stage
$\ct$ and still achieved high entanglement fidelity.
In other words, writing $\rho'\equiv\tr_A\proj{\psi}$ we
have
\beq
F_e(\rho' ,{\cal T} \circ \cd \circ \chi^{\otimes N}) \ge 1-2\epsilon \ .
\label{nop}
\eeq

Achieving a high entanglement fidelity alone is not sufficient.  It is
also necessary to show that $\rho'\equiv {\rm Tr}_{A} \proj{\psi})$
has entropy close enough to that of $\rho \equiv {\rm Tr}_A
\proj{\phi})$ to achieve the same capacity. 
Using Eqs. (\ref{eq37}) and (\ref{part1}) we know that
\beq
|S(\tr_B \proj{\phi}) - S(\tr_B\ \rho^{\rm out})| \le 
2 \sqrt{2 \epsilon} \log \dim \ch_A + 2 \ .
\eeq
for $\epsilon < \frac{1}{72}$.
Since $\tr_B\  \rho^{\rm out} = \tr_B \proj{\psi}$ and
$S(\tr_B \proj{\psi})=S(\tr_A \proj{\psi})=S(\rho')$ and
$S(\tr_B \proj{\phi})=S(\tr_A \proj{\phi})=S(\rho)$ we
have
\beq
|S(\rho)-S(\rho')| \le 2 \sqrt{2 \epsilon} \log \dim \ch_A + 2 \ .
\label{sigmabound}
\eeq
This proves the theorem.  The application to channel capacity is 
straightforward.
As we can always purify a density matrix in a Hilbert space of 
dimension $d$ into a Hilbert space of dimension $d^2$, the dimension
$\dim \ch_A$ can be set to $(\dim \chi)^N$ where
$\dim \chi$ is the dimension on which $\chi$ acts.  
Since the definition of quantum capacity $\qbs$ (\ref{capacityBS}) 
has an $N$ in the denominator, it is clear that (\ref{sigmabound})
strong enough to make $\qbs=\qbs^{\mbox{no encoding}}$.

\section{A correct proof of the capacity of the erasure channel}
\label{sec3}

In this section we will provide a correct upper bound of the capacity
of the erasure channel which \cite{erasure} ``proved'' incorrectly
making use of the unproven assumption that the quantum channel
capacity is continuous.  By providing a correct upper bound
the entire capacity is restored, as the upper bound coincides with
the correct lower bound given in \cite{erasure}.  We work
here with $\qbs$ rather than the definition of capacity in terms
of a protected subspace employed in \cite{erasure} but these
two definitions of capacity have been shown to be equivalent 
\cite{micro1,unpub}.  Cerf independently provided a similar correct upper
bound \cite{cerf} using a slightly different definition of capacity,
which we expect is also equivalent.

Barnum, Nielsen and Schumacher \cite{Barnum} have shown that 
\beq
\qbs^{\mbox{no encoding}} \leq  I_{{\rm max}}^{\mbox{no encoding}}\equiv 
\lim_{N \rightarrow \infty} \max_{\rm \rho} \frac{I_c(\rho,\chi^{\otimes N})}{N}. 
\eeq
Together with the results of Section \ref{sec1} that 
$\qbs=\qbs^{\mbox{no encoding}}$ we now have  
\beq
\qbs \le \lim_{N \rightarrow \infty} \max_{\rho} 
\frac{I_c(\rho,\chi^{\otimes N})}{N}.
\label{fiftyone}
\eeq

A quantum erasure channel with erasure probability $p$
maps an input qubit $\rho$ to $(1-p) \rho + p \proj{3}$ 
where $\ket{3}$ is an orthogonal direction
to the $\ket{1},\ket{2}$ space in which $\rho$ resides.
In \cite{erasure} it was shown correctly that $\qbdsw=0$ for $p \ge 1/2$.
Thus we will here consider only channels with $p < 1/2$.

Recall the definition of the coherent information
\begin{equation}
I_c(\rho,\chi^{\otimes N})=S(\chi^{\otimes N}(\rho))-S_{\rm env}(\rho,
\chi^{\otimes N}).
\end{equation}
For the erasure channel we can write
\beq
I_c(\rho,\chi^{\otimes N})=\sum_{k=0}^N p^k (1-p)^{N-k}
\sum_{i=1}^{{N \choose k}} \left(S(\rho_{i})-S(\rho_{\bar{i}})\right).
\label{icoh}
\eeq
where $i$ designates a particular set of $N-k$ qubits and $\bar{i}$ the complement
of the set $i$. $\rho_{i}$ is defined as $\rho_{i}={\rm Tr}_{\,\bar{i}}\rho$.
This expression is obtained by noticing that the density matrix for the receiver 
is block diagonal where the block labeled with $(i,k)$ is of the form 
\beq
p^k (1-p)^{N-k}\, \rho_{i} 
\eeq
Thus the entropy of the block $(i,k)$ is 
$p^k (1-p)^{N-k}S(\rho_{i})$. 
The total entropy of such a block diagonal density matrix 
$S(\chi^{\otimes N}(\rho))$ is equal to
the sum of the entropy of the blocks plus the entropy of choosing among 
the blocks.
The expression $S_{\rm env}(\rho,\chi^{\otimes N})$ will be the same
as $S(\chi^{\otimes N}(\rho))$ but with $i$ and $\bar{i}$ interchanged 
(what is not erased, the environment gets and vice versa).
Subtracting the two entropies will result in Eq. (\ref{icoh}). 

We split the sum over $k$ into 
two terms, $I_+$ and $I_{-}$, which we will bound separately,
\beq
I_{+}=\sum_{k=0}^{\lfloor N/2\rfloor} p^k (1-p)^{N-k}
\sum_{i=1}^{{N \choose k}} \left(S(\rho_i)-S(\rho_{\bar{i}})\right).
\label{iplus}
\eeq
and 
\beq
I_{-}=\sum_{k=\lfloor N/2 \rfloor +1}^{n} p^k (1-p)^{N-k}
\sum_{i=1}^{{N \choose k}} \left(S(\rho_i)-S(\rho_{\bar{i}})\right).
\eeq
Each term in $I_-$ can be at most
\beq
S(\rho_i)-S(\rho_{\bar{i}}) \leq N-k.
\eeq
To bound $I_{+}$ we will rewrite the sum over the sets $i$ in such a
way that we can use the subadditivity property of the von Neumann entropy.
The idea is to pairwise match terms in Eq. (\ref{iplus}). We match 
$S(\rho_i)$ with a term $S(\rho_{\bar{j}})$ and 
$S(\rho_{\bar{i}})$ with $S(\rho_j)$ where we take the set of qubits $j$ 
such that $\bar{j} \subset i$ and $\bar{i} \subset j$. 
For these matching sets, we can use sub-additivity,
\begin{eqnarray}
S(\rho_i)-S(\rho_{\bar{j}}) \leq N-2k. \\
S(\rho_j)-S(\rho_{\bar{i}}) \leq N-2k
\nonumber
\end{eqnarray}
The way to do the pairwise matching is the following. 
Pick $N-2k$ qubits out of the
total set of $N$ qubits. These are the qubits that two matching sets will have in 
common. Then pick a subset of $k$ qubits out of the remaining $2k$. Together
with the $N-2k$ qubits, these will form set $i$. Set $j$ is made from
the remaining $k$ qubits and the $N-2k$ overlap qubits. In this way each 
set is matched to another one. But we have counted the 
sets multiple times. Each set is counted $2{N-k \choose k}$ times. 
Dividing by this number will thus give us
the original sum. Thus we have derived that
\beq
I_{+} \leq \sum_{k=0}^{\lfloor N/2\rfloor} p^k (1-p)^{N-k}
{N \choose k}(N-2k).
\eeq
We will take $I_{+}$ and $I_{-}$ together and use
\beq
\sum_{k=0}^N {N \choose k}p^k (1-p)^{N-k} k=Np,
\eeq
to get 
\beq
I_c(\rho,\chi^{\otimes N})\leq N(1-p)-
\sum_{k=0}^{\lfloor N/2 \rfloor}{N \choose k}p^k (1-p)^{N-k}k
\eeq
We will use a property of binomial distributions
\begin{eqnarray}
\lim_{N\rightarrow \infty} \frac{1}{N}\sum_{k=0}^{\lfloor N/2 \rfloor}
{N \choose k}p^k (1-p)^{N-k}k=p & \mbox{  for  }p < 1/2.
\end{eqnarray}
This implies 
\beq
\lim_{N\rightarrow \infty} \max_{\rho} \frac{I_c(\rho,\chi^{\otimes N})}{N} \leq  1-2p
\eeq
(note that this bound is achieved by taking $\rho=\ci/2^N$) and therefore 
(with Eq. (\ref{fiftyone}))
\beq
\qbdsw \le \qbs \le 1-2p \  .
\eeq

In \cite{erasure} a constructive lower bound on $\qbdsw$ has been established, 
\beq
\qbdsw \geq 1-2p.
\eeq
Together with our upper bound we prove 
the capacity of the erasure channel
\beq
\qbs=\qbdsw=\max\{1-2p,0\}.
\label{ec}
\eeq

\section{Discussion and Open problems}

An important open question is the conjecture of the equality of $I_{\rm max}$
and the channel capacity. The conjecture would be flawed if $I_{\rm max}
\neq I_{\rm max}^{\mbox{no encoding}}$, since we have shown the 
latter upper bounds the capacity.

Eq. (\ref{ec}) for the capacity of the erasure channel is a continuous
function of $p$, but a resolution of the problem of the continuity of
capacity for general channels is to be desired. If the channel
capacity turns out not to be continuous, this would once again show a
curious characteristic of quantum information.  On the other hand, if
the capacity were proven continuous, the quite general method for
bounding the quantum capacity introduced in \cite{BDSW} and applied
incorrectly in \cite{erasure} would be restored.  For example, the
quantum cloning results in \cite{dagmar} could be used to improve the
bound on the capacity of the quantum depolarizing channel.

In \cite{BDSW} it was shown that the quantum capacities with and without a
classical forward side channel are equal in the case of perfect
error-correction ($\epsilon=0$).  A proof similar to the one in
Sec.~\ref{sec1} can be used to show that this is true for $\qbs$ 
even in the case of asymptotically perfect correction as in the
definition of quantum capacity.  

\section{Acknowledgments} 
The authors would like to thank Charles Bennett, 
David DiVincenzo, and Michael Nielsen for helpful discussions. 
J.A.S. would like to
thank the Army Research Office for financial support. B.M.T. would
like to thank the NWO for financial support from the SIR 
program. H.B. thanks the NSF for financial support under grant
PHY-9722614, and the Institute for Scientific Interchange, Turin, 
Italy, for financial support.

\begin{thebibliography}{99}
\bibitem{superoperator} The most general transformation 
on a density matrix allowed by quantum mechanics is a linear trace-preserving 
completely positive operation. This is equivalent to the unitary operations including an environment initially in a known state $\ket{0}$
as shown in Fig. \protect\ref{fig1}.
\bibitem{kyfan} This is a special case of Ky Fan's inequality, cf. 
\protect\cite{wehrl}.
\end{thebibliography}
\end{document}